\documentclass[manuscript,screen,review=false]{acmart} 

\usepackage{wasysym}
\usepackage{tikz}
\usepackage{graphicx}
\usepackage{subfigure}
\usepackage{subcaption}

\newcommand{\lbl}[1]{\textbf{#1}}

\newcommand{\quo}[1]{``\textit{#1}''}

\AtBeginDocument{%
  \providecommand\BibTeX{{%
    \normalfont B\kern-0.5em{\scshape i\kern-0.25em b}\kern-0.8em\TeX}}}

\copyrightyear{2024} 
\acmYear{2024}

\begin{document}

\title[Students’ Perceptions and Use of GenAI Tools for Programming]{Students’ Perceptions and Use of Generative AI Tools for Programming Across Different Computing Courses}



\author{Hieke Keuning}
\email{h.w.keuning@uu.nl}
\orcid{0000-0001-5778-7519}

\author{Isaac Alpizar-Chacon}
\email{i.alpizarchacon@uu.nl}
\orcid{0000-0002-6931-9787}

\author{Ioanna Lykourentzou}
\email{i.lykourentzou@uu.nl}
\orcid{0000-0002-4243-4128}

\author{Lauren Beehler}
\orcid{0009-0008-2352-3935}
\email{l.a.beehler@uu.nl}

\author{Christian Köppe}
\email{c.koppe@uu.nl}
\orcid{0000-0003-0326-678X}

\author{Imke de Jong}
\email{i.dejong1@uu.nl}
\orcid{0000-0003-0404-4011}

\author{Sergey Sosnovsky}
\email{s.a.sosnovsky@uu.nl}
\orcid{0000-0001-8023-1770}

\affiliation{%
  \institution{Utrecht University}
  \city{Utrecht}
  \country{The Netherlands}
}

\renewcommand{\shortauthors}{Hieke Keuning, Isaac Alpizar-Chacon, Ioanna Lykourentzou, et al.}

\begin{abstract}

Investigation of students' perceptions and opinions on the use of generative artificial intelligence (GenAI) in education is a topic gaining much interest. Studies addressing this are typically conducted with large heterogeneous groups, at one moment in time. 
However, how students perceive and use GenAI tools can potentially depend on many factors,  including their background knowledge, familiarity with the tools, and the learning goals and policies of the courses they are taking. 

In this study we explore how students following computing courses use GenAI for programming-related tasks \textit{across different programs and courses}: Bachelor and Master, in courses in which learning programming is the learning goal, courses that require programming as a means to achieve another goal, and in courses in which programming is optional, but can be useful.
We are also interested in changes over time, since GenAI capabilities are changing at a fast pace, and users are adopting GenAI increasingly.

We conducted three consecutive surveys (fall `23, winter `23, and spring `24) among students of all computing programs of a large European research university. We asked questions on the use in education, ethics, and job prospects, and we included specific questions on the (dis)allowed use of GenAI tools in the courses they were taking at the time.

We received 264 responses, which we quantitatively and qualitatively analyzed, to find out how students have employed GenAI tools across 59 different computing courses, and whether the opinion of an average student about these tools evolves over time. Our study contributes to the emerging discussion of how to differentiate GenAI use across different courses, and how to align its use with the learning goals of a computing course. 

\end{abstract}



\maketitle

\section{Introduction}

An increasing number of studies have investigated students' perceptions and opinions on the use of generative artificial intelligence (GenAI) within education in general, and specifically for computing students (e.g. \cite{prather2023robots, smith2024early,rogers2024attitudes}). These studies are typically conducted either by inviting a large  body of computing students to complete a survey once, or within a specific computing course. 
How generative AI is used by students in their studies might change depending on the learning goals of the course they are taking. Students taking an introductory programming course might use GenAI for explaining compiler error messages, or explaining a programming concept. They would not learn much from generating the solution for each programming exercise, but some of them probably do that anyway, for different reasons. In follow-up courses this might be different; if the learning goal is to build a website using a framework, generating small snippets for subtasks might be acceptable and not impede their learning.
In this study we explore further how students use and perceive GenAI across different programs and courses: Bachelor and Master, courses in which learning programming is the learning goal, and courses in which programming is a means to achieve another goal. We also investigate changes over time, since GenAI capabilities are changing at a fast pace, and users are adopting GenAI increasingly.
We conducted three consecutive surveys among students of all programs of the Information and Computer Science department of a large European research university, capturing their perceptions on job prospects, policies and ethics, and classroom use. The survey also included a course-specific part to investigate to what extent students use it for programming-related tasks, and their beliefs on whether that should be allowed, or not.
We conducted a quantitative and qualitative analysis to answer the following research questions:

\textbf{RQ1}: How are GenAI tools currently used by students across different types of programming courses?

\textbf{RQ2}: How have students' opinions on GenAI changed over the course of an academic year? 

The contributions of this paper are: 1) an updated snapshot of students' perspectives from the academic year 2023-24, 2) an overview of changes over time, and 3) a fine-grained analysis of GenAI use cases and rationale for programming-related tasks within different computing courses. 
The insights provide a background to the increasing adoption of GenAI by students, and the emerging discussion of how to differentiate generative AI use across different courses, and how to align its use with the learning goals of computing courses. 

\section{Background}
\label{sec:bg}

Since the first paper that discussed GenAI's implications for computing education \cite{finnie2022robots}, many studies have appeared. While the first papers mainly focused on GPT's and Codex's abilities to generate solutions to standard programming problems, the topics expanded to educational content generation \cite{sarsa2022automatic}, student's use of GenAI in the classroom \cite{kazemitabaar2023studying,hellas2024experiences,prasad2023generating,margulieux2024self}, and its implications for computing education \cite{becker2023programming,becker2023generative,prather2023robots}.
Some of these studies look at the introduction of GenAI from the perspective of an educator. \citeauthor{becker2023programming}, for example, identified several challenges and opportunities related to introductory programming courses in winter 2022-23 \cite{becker2023programming}. The challenges that were identified relate (among others) to the risk of over-reliance by students on tools (impeding learning of fundamental concepts or problem solving ability), and academic misconduct in the re-use of code. Other studies have found these proposed challenges to be realistic. \citeauthor{gooch2024exploring} \cite{gooch2024exploring} analyzed a large number of university student assessments in the summer of 2022 and 2023, and noticed an increase in material flagged as AI-generated. Although the authors had expected a larger percentage to be AI-flagged, it clearly shows that students are using GenAI tools actively.
Next to the challenges discussed earlier, \citeauthor{becker2023programming} \cite{becker2023programming} also identified  opportunities for programming education, including the ability to quickly generate example solutions to problems, support in reviewing of students' code, and generating exercises.  Examining the opportunities for enriching education more closely, \citeauthor{cambaz2024use} \cite{cambaz2024use} conducted a systematic literature review, identifying teaching and learning practices that use GenAI. The learning practices they found are: `generate practice exercises', `generate exemplar solutions', `generate alternative solutions', `improve student code', `clarify error messages and provide suggestions', `support conceptual understanding, `provide syntax tips', and `code explanation'. In the context of software engineering education, \citeauthor{daun2023chatgpt} \cite{daun2023chatgpt} discuss that GenAI can offer numerous benefits, as long as students are supported in the use of these tools.

Other studies investigated \textit{student's use} of GenAI.
\citeauthor{prather2024weird} \cite{prather2024weird} conducted the first study of computing students working with Copilot, identifying two novel interaction patterns: `drifting' when students are lost, accepting and working with incorrect suggestions, and `shepherding' in which students try to force the tool to generate code they think they need.
In other work, \citeauthor{prather2024widening} \cite{prather2024widening} conducted a think-aloud and eye-tracking study with 21 students working on a programming exercise with Copilot and ChatGPT. The authors identified several meta-cognitive difficulties, finding an increase in the divide between students who would be doing well anyway, and struggling students. High-scoring students used GenAI to amplify their work, but for low-scoring students GenAI tools introduced additional difficulties, which they were often not even aware of. They also found that students sometimes not even realize that they are using AI tools, such as accepting some of Copilot's continuous suggestions. 
\citeauthor{margulieux2024self} \cite{margulieux2024self} studied student's use of GenAI, focusing on the relationship between GenAI use and self-regulation strategies, self-efficacy, and their fear to fail at programming. They conducted a study with 40 students, collecting data multiple times during an introductory programming course. They found that some students use GenAI to their benefit, still wanting to learn rather than over-relying on it. Students with higher self-efficacy, lower fear of failure, or higher prior grades used GenAI less, found it less useful, and used it later in their problem-solving.

Several studies focus on \textit{students' views} on the emergence of Generative AI as a new tool they could use in their computing studies.
\citeauthor{prather2023robots} \cite{prather2023robots} conducted a survey among 171 computing students from 17 countries (with the top three being New Zealand, Jordan, and the USA) in the summer of 2023, as part of an ITiCSE working group on Generative AI in computing education. As far as we know, this was the first survey among students in the computing education context specifically. They found that at that time, the use of GenAI tools was still limited. Also, opinions among students and teachers (which were also surveyed) were pretty much in line, both agreeing that there should be restrictions in place for GenAI use.
\citeauthor{smith2024early} \cite{smith2024early} conducted a survey in spring 2023, shortly after ChatGPT's release, with the aim to capture a snapshot of students' views at a time in which little was known about the consequences, and policies were not in place yet.
They surveyed 133 CS Majors (of which 116 undergraduate) at a small engineering-focused research university in the US.
They found that most students had used GenAI tools for various tasks, and leaned towards a positive view.

\citeauthor{cipriano2024chatgpt} \cite{cipriano2024chatgpt} surveyed 52 CS students (all first-years), also in early 2023. In a course on data-structures and algorithms, students were given an assignment in which using ChatGPT was explicitly required. Students were asked about their opinions on and use of ChatGPT for academic purposes in a survey administered at the end of the course. The results show that most students (44) were in favor of teachers allowing the use of ChatGPT for assignments, and all students agreed that (ungraded) assignments require using tools like ChatGPT should be included in the curriculum.
In another study done not long after the introduction of ChatGPT, \citeauthor{rogers2024attitudes} \cite{rogers2024attitudes} surveyed 70 US students (CS majors and minors) at one institution about their use of and attitude towards using the tool. The researchers also asked students for what kinds of tasks they think ChatGPT should or should not be allowed. Most students agreed it is okay to use the tool for brainstorming or finding references. About a third of the students (32.4\%) agreed it should be allowed to use it for writing a first draft of a paper. The majority did agree ChatGPT should be credited if used, and that it should not be used for the final draft of a paper. \citeauthor{rogers2024attitudes} \cite{rogers2024attitudes} also inquired for what purposes students were using ChatGPT. The results showed it was used mainly to help in understanding CS concepts and studying for exams.

There are other studies that examined what kinds of tasks students use GenAI tools for. In a study conducted by \citeauthor{budhiraja2024jarvis} at multiple universities in India, ChatGPT usage was investigated through a survey and interviews \cite{budhiraja2024jarvis}. A total of 480 undergraduate students in Computer Science responded to the survey. The results showed students mostly used the tool for gathering information, summarizing content, and assistance in coding. Most students were positive about the use of ChatGPT as a learning tool, as it can facilitate personalized learning, give immediate feedback and provide additional information where necessary. The 17 interviews that were conducted support these results, but also underscored that students do no always trust the solutions of the tool to be correct, leading to a lot of effort spent in correcting the output.
In a survey of 253 students, \citeauthor{amoozadeh2024trust} \cite{amoozadeh2024trust} examined the use of GenAI tools, and also asked questions about the level of trust students place in the results of these tools. Undergraduate and graduate students at two institutions in India and the US were asked about their experience with GenAI for both programming and non-programming tasks. The results showed that students use the tools for both, although for programming mainly to seek help in understanding code, not for generating new code. The authors conclude GenAI tools may replace internet searches as a way of finding help. The survey also showed that most students do not blindly trust that output generated by AI tools is always correct.
Previously mentioned studies found GenAI to provide a new avenue for information seeking. \citeauthor{hou2024effects} \cite{hou2024effects} looked more closely at this type of usage. Through a survey answered by 47 computing students across multiple universities in the US, they found online searches to still be the most predominantly used channel for information seeking on an hourly or daily basis. ChatGPT was used less frequently, although still at least weekly by 42.6\% of the students. They also asked students about their comfort levels consulting peers, TA's and teachers or GenAI for help, and found that students were more comfortable asking ChatGPT for help than asking TA's and teachers publicly. Interviews with 8 participants revealed for what information seeking needs the students preferred to use ChatGPT and why. For example, it was found convenient that you could keep asking questions when concepts were unclear, without concerns of being a bother to someone. This was also mentioned in the context of coding, where students found it helpful that tools like ChatGPT can suggest multiple ways to approach a problem.

The studies described above provide different snapshots of the use of GenAI tools by students since the introduction of ChatGPT in late 2022. They often ask students about their use of tools \textit{without} the context of a specific course. With the current study we aim to add to this body of knowledge in two ways. First, we believe an examination of changes in perspectives and use of these tools over time would be valuable. As GenAI applications become more commonplace and the tools are further developed, one may expect developments in the perspectives and use of these tools as well. Second, we are also interested in examining students' perspectives on the use of these tools for different types of courses. In the context of CS education, some courses aim to teach students programming skills, while other courses require them to put acquired skills to use to complete projects. It is possible that students use the tools differently for different types of courses. For example, students might use tools to clarify concepts or code constructs when still learning to program, but to generate initial code snippets when they know who to program (and are thus able to correct any errors made in the generated code). 
With the current study, we aim to provide this longitudinal and course type differentiated perspective.

\section{Method}

We conducted a survey among students of the department of Information and Computing Sciences (ICS) at Utrecht University. The survey was distributed at three instances during the academic year 2023--2024: after Period 1 (November 2023), 2 (February 2024), and 3 (May 2024). The academic year at Utrecht University is divided into two semesters, having two periods each. 

Utrecht University is a large research university in the Netherlands. The ICS department offers Bachelor (BSc) and Master’s (MSc) programs to over 2,000 students. These programs cover disciplines such as Computing Science, Information Science, Game and Media Technology, Data Science, Human-Computer Interaction, Business Informatics, and AI. The Bachelor program and its courses are mostly offered in Dutch, and the Master programs are in English with many students from abroad attending. Courses are typically 7.5 European Credits, and many students would take two of those courses in one period. Several courses are also open to students from outside the department.

\subsection{Survey design}

The survey was largely based on the survey conducted as part of the 2023 ITiCSE working group on Generative AI in computing education \cite{prather2023robots}, as mentioned in Section \ref{sec:bg}.
We made some minor adjustments to this survey, e.g. we included our BSc and MSc programs, removed a few questions we considered irrelevant in the context of the current study, and improved some phrasings. However, we kept most questions to be able to compare results. We added some questions about the students' (perceived) programming proficiency (Q6) and importance for their future career (Q7).
We divided the survey questions into two parts: part 1 (general) contained demographics and general questions on student's perceptions of GenAI, and part 2 (course-specific) contained questions on students' GenAI use and rationale for that, from which we expect the answers will vary for different courses. At the end of part 1 of the survey, we added a question on which courses the student had followed in the period before. For each selected course, the survey then dynamically offered the set of course-specific questions.

Because we conducted the survey multiple times, some students might participate more than once. To prevent students from abandoning the survey when confronted with a large set of questions they answered before, we included a question about former participation. If they stated they did the survey before, they would skip the general part of the survey and go directly to the course-specific questions.

The participants were asked for consent to join the survey, and to agree that the results (including anonymized quotes) may be used and shared in academic and non-academic publications. 
The consent form also states that participation is voluntary, participants have the right to withdraw at any point, responses will be kept completely confidential, stored on a secure server, and data will be deleted 5 years after the conclusion of the project. We also gave contact information of the education director and privacy officer, who were not involved in the study.

The final survey can be found in the supplemental materials. The survey was created using Qualtrics, hosted on a university server, and the data was later stored on a protected research data drive. 

\subsection{Distribution}

For each period, we assembled a list of the courses offered within the study programs of the ICS department. We excluded thesis courses and supporting courses with very little credits (e.g. seminars). We labeled each course regarding the role of programming in the course. We focused on textual languages that can be easily generated by GenAI, therefore including languages such as SQL, CSS, and HTML. We distinguish between the following types of courses:
\begin{itemize}
    \item \textsc{ProgCrs.} Programming course. These are core programming courses, in which learning to program is the major learning goal. Examples are `imperative programming' and `advanced functional programming'. 
    \item \textsc{ProgReq.} Although not a learning goal in itself, programming is required for this course. Examples are software development project (capstone) courses, computer graphics, AI courses, algorithms, etc.
    \item \textsc{ProgOpt.} Programming is optional. Programming knowledge is not needed, but can be used (by students who want to) in the course, for example to do some data analysis in Python. This category also includes courses in which programming is needed for a group project, but not all students have to contribute to the coding part. Lastly, this category includes a course using a low-code platform, since some programming could be useful there.
    \item \textsc{NoProg.} No programming component. Examples of these are courses about research methods, or ethics within computing. 
\end{itemize}
We asked teachers to choose the right label for their own course. If we did not receive a reply, we looked at the course description and used our own knowledge of courses to determine the label. We include the final list of courses for which we received responses in the supplemental materials.

At the end of each period, we sent a request to the teachers of the courses for that period (except those without any programming component), asking them to distribute a message to students in their course to participate in the survey. Teachers could use email, the learning management system, or another channel used in the course to invite the students. We decided to distribute the survey in the context of a course, because of the more personal approach. A downside of this was that not all teachers acted upon our request. However, because students could select all courses they had followed in the survey, we could still get some data from courses through which the survey was not distributed.

For each survey, we randomly selected three participants to win a € 15 gift voucher. They had to leave an email address if they wanted to join this raffle. These addresses were collected in a separate part at the end of the survey, not connected to their responses.

\subsection{Data analysis}

We cleaned the received survey data by removing responses that only contained answers to the demographic questions, and some repeated responses flagged as invalid (ballot stuffing).
We added a course type label for each course-specific response, distinguishing between BSc and MSc courses, resulting in 8 groups.
We updated a few course types only after the survey to NoProg, which we decided to keep. 
Based on the number of responses for each of the 8 groups, we merged groups with few responses, resulting in 5 new groups: \textsc{ProgCrs-BSc}, \textsc{ProgReq-BSc}, \textsc{ProgReq-MSc} (merged with ProgCrs-MSc), \textsc{ProgOpt-BSc} (merged with NoProg-BSc), and \textsc{ProgOpt-MSc} (merged with NoProg-MSc).

For this paper, we do not present an exhaustive descriptive analysis of all survey questions. Instead, we focus on the questions that contribute to answering our research questions. 

\subsubsection{Quantitative analysis}

We analyzed the responses to all closed questions from the general part of the survey (Q8 - Q13) over the three periods. For each question, we conducted a temporal analysis to observe changes and trends over time. To facilitate a clear and comprehensive visualization, we created plots depicting the responses across the three periods. Given that the majority of our data is ordinal, derived from Likert-scale questions, we employed the Mann-Whitney U test to determine if there were statistically significant differences between the periods.

For the course-specific section, we focused on two questions (Q17 and Q19). In addition to analyzing these questions over time, we examined the data across different course levels (BSc and MSc) and the five course types. We applied both the Mann-Whitney U test and the Kruskal-Wallis H test to account for more than two groups. Furthermore, we utilized Python as a supporting tool for the analysis.

\subsubsection{Qualitative analysis}

We analyzed the responses to two open questions that the students had to answer for each course they had taken, namely ``Describe the ways you currently use GenAI tools in this course for code generation (e.g.: debugging, writing, etc.)?'' (Q22) and ``Can you elaborate on when you believe GenAI should be allowed or disallowed in this course?'' (Q20). These questions contained 269 and 284 student answers respectively. To analyze these, we followed the principles of Design Thinking, employing affinity diagramming to organize and categorize the qualitative data. Design thinking was chosen as a human-centered, collaborative, and iterative method that is particularly effective in synthesizing diverse perspectives and fostering a deeper understanding of participant answers, making it a suitable approach for conducting thematic analysis to uncover hidden patterns in the qualitative data~\cite{braun2022conceptual, brown2009change}.

For Q20, two authors began by extracting key quotes from the user responses and then grouping similar quotes together. Responses were color-coded according to the type of course. This process allowed us to identify common themes and patterns across the answers. Next, we engaged in pattern synthesis to combine these insights, creating a cohesive understanding of the student perspectives. By continuously refining and re-evaluating these groupings, we were able to uncover deeper insights and more nuanced trends. Finally, we combined these insights to form a holistic view.
This iterative and collaborative approach ensured that we captured the richness of the qualitative data and translated it into insights that could effectively complement the quantitative analysis. The Miro board software\footnote{\url{https://miro.com/}} was employed for this process, as it provides a user-friendly visual environment that facilitates real-time collaboration, enabling us to process a large set of color-coded responses efficiently. While more specialized software exists, offering advanced features for qualitative analysis, Miro was chosen for its flexibility and ease of use, allowing for dynamic grouping of the quotes in a way that enhanced both the transparency and collaborative nature of the analysis. 

For Q22, we used an existing taxonomy from \citeauthor{tufano2024unveiling}~\cite{tufano2024unveiling} for types of software development tasks that can be automated using ChatGPT as a starting point for the analysis. This taxonomy was developed based on a large number of commits, pull requests, and issues, which were mined from software repositories. We considered using the codebook developed by \citeauthor{smith2024early} \cite{smith2024early}, but theirs had only a few coding-related tasks. Two authors first grouped the responses according to the top-level categories in \citeauthor{tufano2024unveiling}'s taxonomy. In the next round, we further refined by assigning responses to subcategories and developing new subcategories that emerged from the responses.

\section{Results}

Table \ref{tab:responses} gives an overview of the number of responses per period. In total, we got 264 responses to the overall survey. The general section received 249 responses, while the course-specific section received 340. Respondents included 47 females, 206 males, 1 non-binary individual, 2 participants who self-described, and 8 who did not indicate their gender. Of the responses, 163 were from BSc students (across 3 programs from the ICS department and 9 programs from other departments), 99 from MSc students (across 7 programs from the ICS department and 3 programs from other departments), and 2 from PhD students. Students have a diverse range of (self-estimated) programming proficiency, with BSc students averaging 6.4 out of 10, and MSc students averaging 7.3.
The survey also shows that BSc students place high importance on programming for their future careers, rating it on average 7.9 out of 10. MSc students have a slightly higher perception, with an average rating of 8.2.

\begin{table}[bt]
\centering
\caption{Overview of the number of responses.}
\label{tab:responses}
\sffamily
\begin{tabular}{lrrrr}
\hline
 & Period 1 & Period 2 & Period 3 & \textbf{Total} \\ \hline
Responses to general part   & 139 & 56 & 54 & 249 \\ 
Responses to course part    & 155 & 83 & 102 & 340 \\ 
Different number of courses & 19  & 20 & 22 & 59* \\ \hline
\end{tabular}
\vspace{1mm} 
\parbox{.6\linewidth}{\footnotesize * One course spans all periods.}
\end{table}

\begin{figure}[bt]
    \centering
    \subfigure[I regularly use GenAI when working with code (Q8b)]{\includegraphics[width=0.3\textwidth]{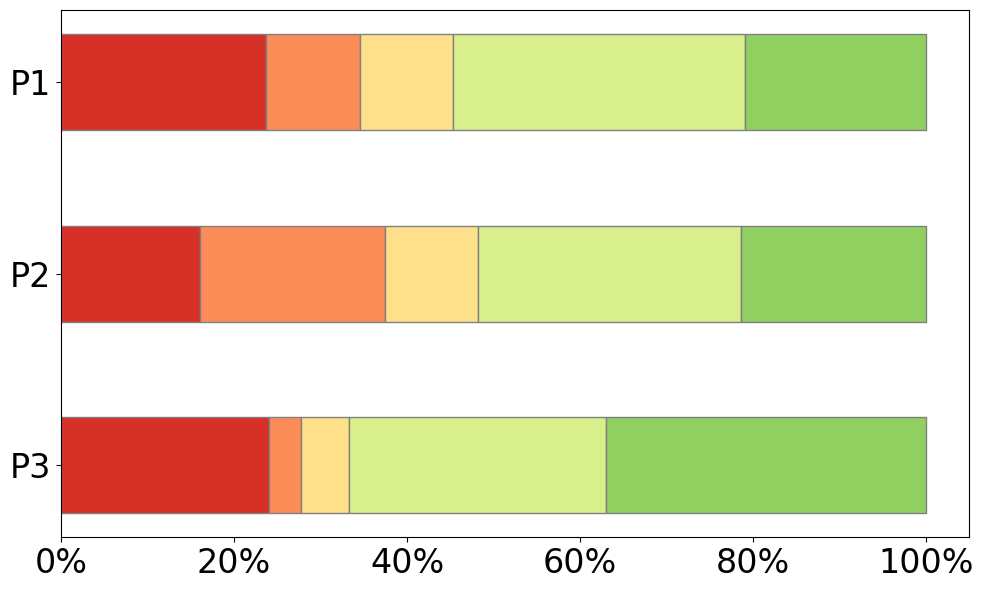}\label{fig:q8b}}
    \subfigure[I expect to use GenAI for my learning practices in the future (Q9a)]{\includegraphics[width=0.3\textwidth]{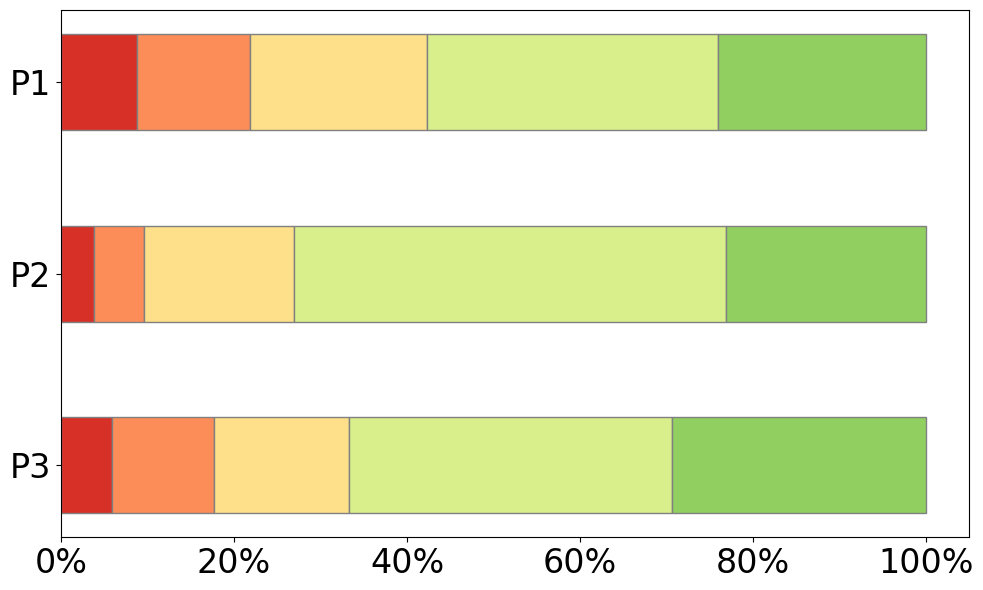}\label{fig:q9a}}
    \subfigure[Using GenAI frequently is harmful for my learning of programming (Q9b)]{\includegraphics[width=0.3\textwidth]{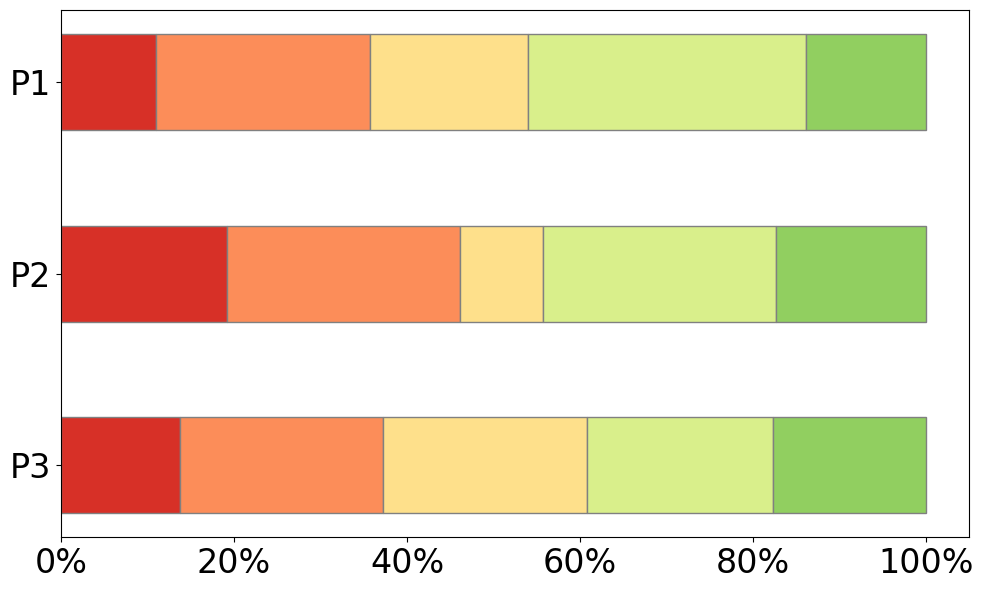}\label{fig:q9b}}
    \subfigure[GenAI will negatively impact my future job prospects (Q13a)]{\includegraphics[width=0.3\textwidth]{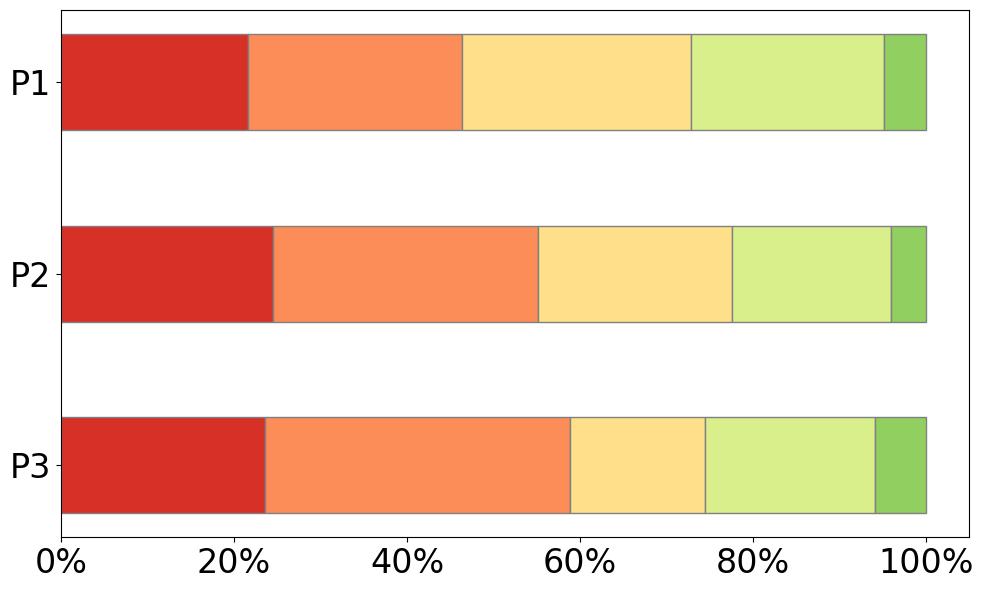}\label{fig:q13a}}
    \subfigure[I am concerned that I will become overreliant on GenAI tools (Q13c)]{\includegraphics[width=0.3\textwidth]{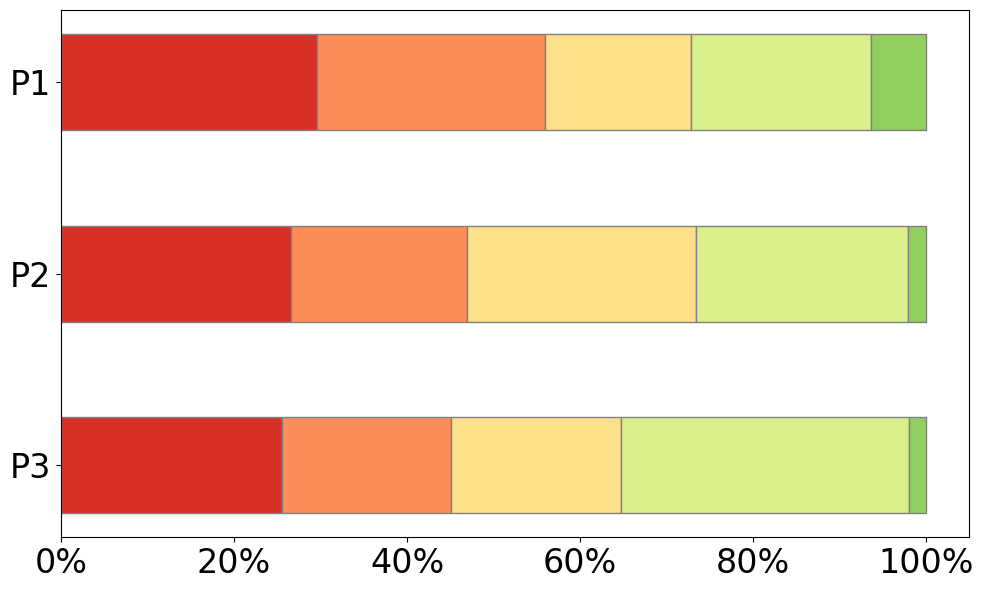}\label{fig:q13c}}
    \subfigure[I trust the code written by GenAI more than the code I write (Q13d)]
    {\includegraphics[width=0.3\textwidth]{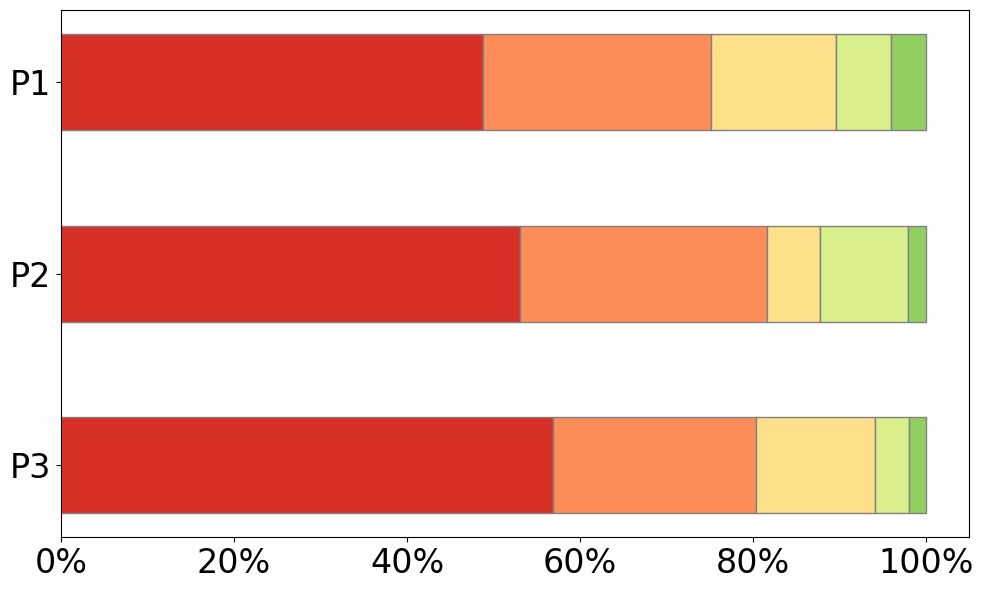}\label{fig:q13d}}
    {\includegraphics[width=0.85\textwidth]{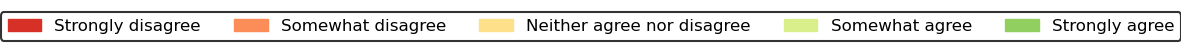}}
    \caption{Opinions on (future) usage and concerns of GenAI over time.}
    \label{fig:grid}
\end{figure}

\begin{figure}[ht]
    \centering
    \subfigure[Students’ unethical use of GenAI (Q11)]{\includegraphics[width=0.45\textwidth]{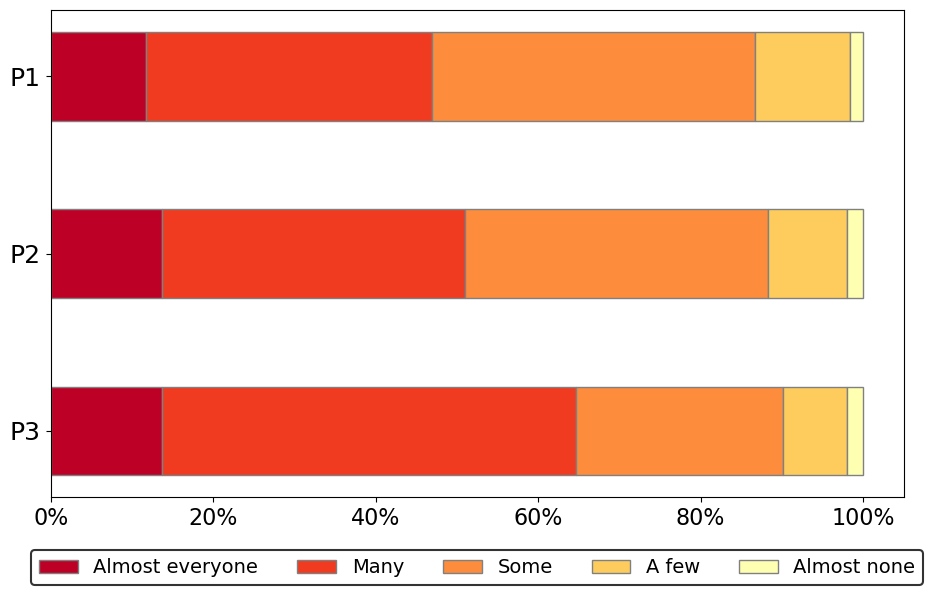}\label{fig:q11}}
    \subfigure[Ethical and unethical uses of GenAI (Q12d\&a)]{\includegraphics[width=0.5\textwidth]{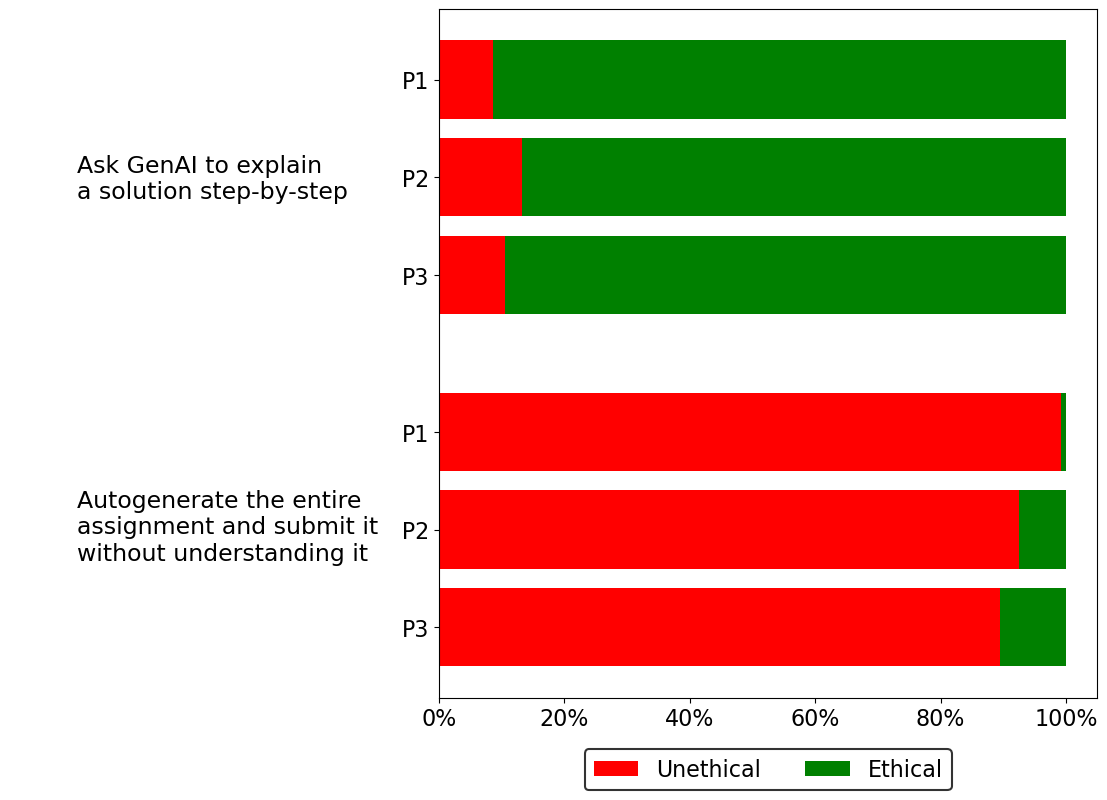}\label{fig:q12}}
    \caption{Opinions on ethics of GenAI.}
    \label{fig:grid_2}
\end{figure}

\subsection{General impressions and trends}\label{sec:results_general}
Overall, we observe small variations in student perceptions and usage over time for the general questions. However, these differences are not statistically significant, with a few exceptions. Figures \ref{fig:grid} and \ref{fig:grid_2} display the responses to select questions over the three periods, representing three key themes of the survey: (future) usage, concerns, and ethics.

In terms of usage, students in Period 3 report a higher agreement (somewhat and strongly) with the statement that they use GenAI regularly when working with code, compared to previous periods (Figure \ref{fig:q8b}). Furthermore, a slightly higher number of students in Period 3 agree more strongly that they will use GenAI for their future learning than in the other periods (Figure \ref{fig:q9a}). Regarding concerns, students' perceptions of the negative impact of GenAI on their programming learning (Figure \ref{fig:q9b}) and future job prospects (Figure \ref{fig:q13a}) have remained stable. However, students in Period 3 express more concern (somewhat and strongly agree) about becoming over-reliant on GenAI (Figure \ref{fig:q13c}). This trend in usage and concerns about over-reliance appears to follow the increase in popularity and enhanced capabilities of these tools, particularly with the release of ChatGPT-4 and its updated version, ChatGPT-4o (May 2024\footnote{\url{https://openai.com/index/gpt-4o-and-more-tools-to-chatgpt-free/}}). However, students continue to trust their own code more than the code written by GenAI (Figure \ref{fig:q13d}). Although newer models are more capable of coding tasks \cite{prather2023robots,coello2024effectiveness}, their perception of trust has not changed over time. One possible explanation is that students predominantly use older and free versions, such as ChatGPT-3.5 (ChatGPT-4o was released just at the end of the last iteration of the survey).

Students' opinions related to the ethical aspects of GenAI have shown some interesting changes. In Period 3, there was an increase in the belief that more students are using GenAI tools in ways that their instructors would not approve (Figure \ref{fig:q11}). Specifically, the "many" category increased, while the "some" category decreased, compared to Periods 1 and 2 (the other categories remained similar). However, a Mann-Whitney test did not show statistical significance (Period 3 vs. Period 1, \( U = 2740 \), \( p = 0.076 \)). In Q12, we asked students to consider whether different situations were ethical or not. Figure \ref{fig:q12} shows two very different cases. In the first case (top of the figure), the majority of students considered that using GenAI to explain a solution step-by-step is ethical. A Chi-Square test confirmed that this perception has remained unchanged throughout the three periods (\( \chi^2(2) = 0.90 \), \( p = 0.639 \)). In the second case (bottom of the figure), there is a noticeable change in the number of students who considered it ethical to auto-generate an entire assignment and submit it without understanding it. While only 1\% of students considered this ethical in Period 1, this number rose to nearly 8\% in Period 2 and further increased to almost 11\% in Period 3. This clear upward trend was confirmed with a Chi-Square test (\( \chi^2(2) = 9.82 \), \( p = 0.007 \)). Although we do not have data that explicitly explains this trend, it may be attributed to the rapid adoption of GenAI by students and a lack of education on the ethics of AI, as well as the potential failure of instructors to modify or adapt their assignments to address GenAI usage.

\subsection{Perceiving GenAI as a source of support}\label{sec:results_q17}

Q17 asked students to rank six potential sources of support according to the order in which they would resort to them in the case of a problem. These sources were: GenAI, Course discussion forum, Online search, Friend, Online forum (e.g. StackOverflow), and Course teacher or a TA. This question is particularly interesting, as we can estimate the role GenAI tools now play in computing courses compared to all other traditional practices of help seeking. Figure \ref{fig:two_images} presents the results of the two analyses we conducted over the 340 responses students provided to this question.

\subsubsection{Trend over time}

\begin{figure}[h]
    \centering
    \subfigure[Trends over time]{\includegraphics[width=0.45\textwidth]{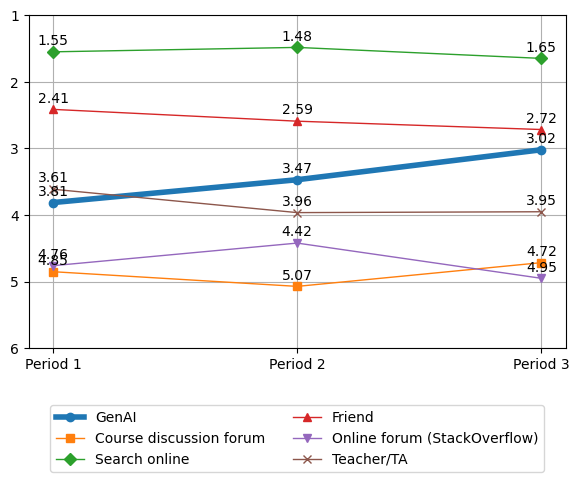} \label{fig:q17_overtime}}
    \subfigure[Trends by course type]{\includegraphics[width=0.45\textwidth]{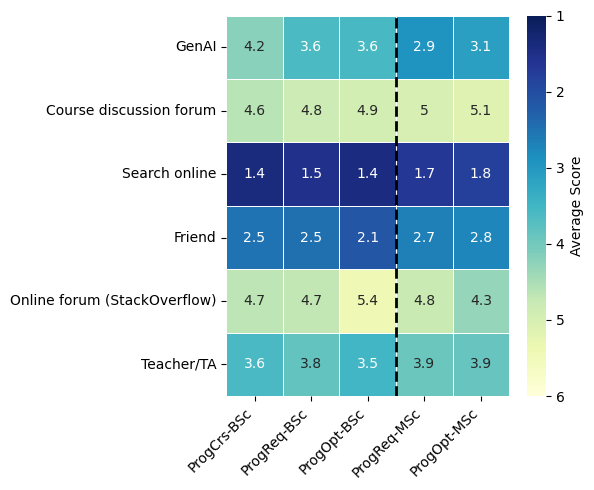} \label{fig:q17_hm}}
    \caption{Average prioritization of different help-seeking strategies when students encountered difficulties in the course (Q17). 1 is more important, 6 is less important.}
    \label{fig:two_images}
\end{figure}

First, we wanted to see if students' opinions regarding the usefulness of GenAI as a learning tool changes over time. Figure \ref{fig:q17_overtime} shows the effect of this progress on the students' perception of GenAI. Over three terms, it moved from the close forth to the close third choice, surpassing teacher/TA in the process. In fact, GenAI is the only source of help that has been consistently gaining prominence at the expense of other sources. This is not surprising; as the GenAI tools improved and more students learned about them and learned how to use them, their popularity grew. 

\subsubsection{Group analysis}

We have also looked at how differently students rank GenAI as a support tool / help-seeking strategy across different course types (see Figue \ref{fig:q17_hm}). This analysis has revealed two main results. When comparing MSc versus BSc courses, we have found that MSc students rank GenAI significantly higher as a support tool than BSc students do; the results of the Mann-Whitney test for course level are \( U = 9459.5 \), \( p < 0.001 \). Looking at different course categories with each level, we did not see any difference for MSc-level courses, however, we did observe a significant difference on the BSc-level. Students of the introductory programming courses (ProgCrs-BSc) prioritized GenAI tools  significantly lower as a help-seeking strategy than students of the BSc courses where programming was a required (ProgReq-BSc) or an optional (ProgOpt-BSc) component. The results of the Kruskal-Wallis test for course category on the BSc-level are as follows \( \chi^2(2) = 6.20 \), \( p = 0.045 \). These two findings are consistent with each other. BSc students are less proficient in programming than MSc students, and  BSc students taking introductory programming are the least proficient of all. As students' programming expertise grows, they are more ready to engage with GenAI tools.

\subsection{Acceptance of GenAI for programming assignments}
For each course the students took, we asked if they believed GenAI should be allowed or disallowed for programming assignments (Q19). Students could choose from three options: always allowed, allowed in some assignments but disallowed in others (mixed acceptance), or always disallowed. Figure \ref{fig:q19} presents the results of our analysis based on the 340 responses to this question.

\subsubsection{Trend over time} We aimed to investigate whether the acceptance of GenAI for programming assignments has changed over time in the BSc and MSc courses. Figures \ref{fig:q19_time_bsc} and \ref{fig:q19_time_msc} illustrate the acceptance trends over time for each group. At the BSc level, the distribution appears stable across periods. This observed stability was confirmed with a Kruskal-Wallis test (\( \chi^2(2) = 1.53 \), \( p = 0.465 \)). In contrast, for the MSc level, there is an increasing trend over the periods. However, the results of the Kruskal-Wallis test indicate that these changes are not statistically significant (\( \chi^2(2) = 5.12 \), \( p = 0.077 \)). These findings suggest that despite the increasing use of GenAI among students (as noted in Q8---see Section \ref{sec:results_general} and Q17---see Section \ref{sec:results_q17}), their views on its use for assignments have remained relatively unchanged. Furthermore, a Kruskal-Wallis test comparing the time distributions of both groups reveals significant differences between BSc and MSc students (\( \chi^2(5) = 27.16 \), \( p < 0.0001 \)). This shows that students in MSc courses are more open to using GenAI in programming assignments compared to BSc courses.

\subsubsection{Group analysis} 
\begin{figure}[t!]
    \centering
    \subfigure[Over time in BSc courses]{\includegraphics[width=0.43\textwidth]{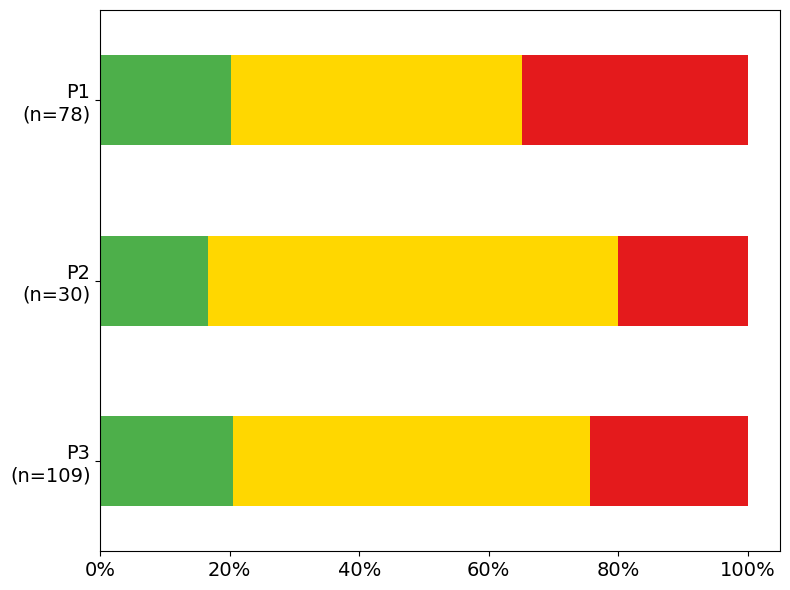} \label{fig:q19_time_bsc}} 
    \subfigure[Over time in MSc courses]{\includegraphics[width=0.43\textwidth]{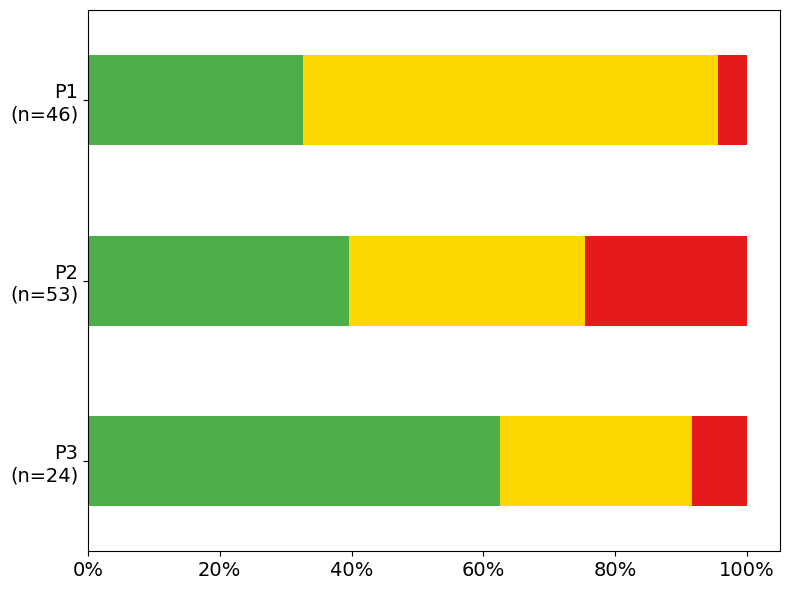}\label{fig:q19_time_msc}}
    \subfigure[Grouped by course level]{\includegraphics[width=0.43\textwidth]{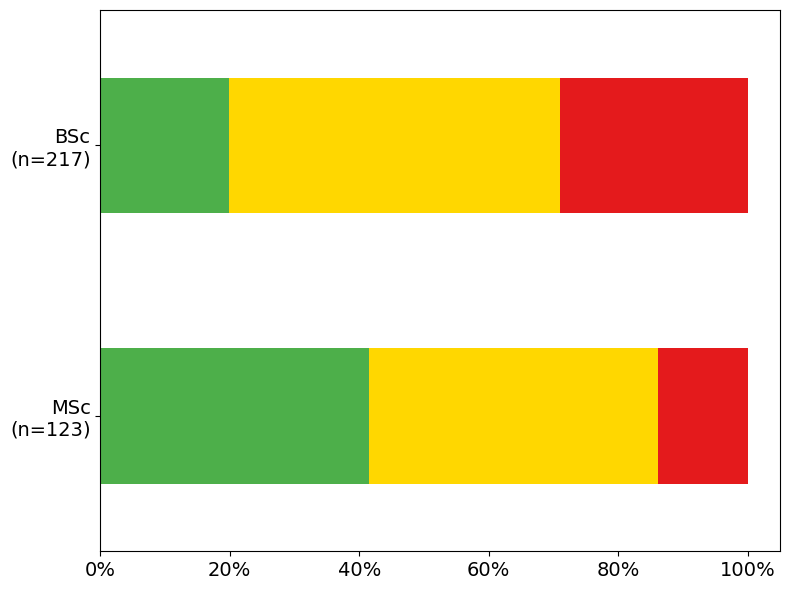}\label{fig:q19_level}}
    \subfigure[Grouped by course type]{\includegraphics[width=0.43\textwidth]{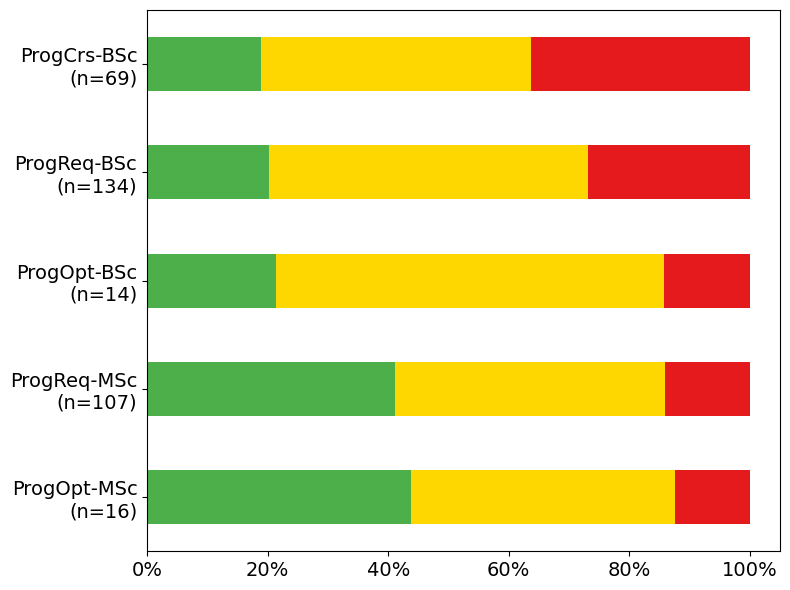}\label{fig:q19_course_type}}  \\
    {\includegraphics[width=0.4\textwidth]{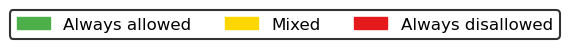}}
    \caption{Opinions on (dis)allowing GenAI for programming assignments (Q19).}
    \label{fig:q19}
\end{figure}

The observed difference between MSc and BSc students becomes more evident when we aggregate the responses by course level and course type (Figures \ref{fig:q19_level} \& \ref{fig:q19_course_type}). Consistent with our earlier findings, MSc students generally exhibit a higher acceptance of GenAI usage compared to BSc students. Specifically, a larger proportion of MSc students support allowing GenAI either always or in a mixed manner, while a significant portion of BSc students prefer to disallow its use. This trend persists across different types of courses, indicating a broader acceptance of GenAI among more advanced students. Statistical tests confirmed this difference: a Mann-Whitney test for course level (\( U = 9667.5 \), \( p < 0.0001 \)) and a Kruskal-Wallis test for course type (\( \chi^2(4) = 22.76 \), \( p < 0.001 \)).

\subsection{Using GenAI for programming across different courses}

In this section we dive deeper into \textit{how} and \textit{why} students (do not) use GenAI tools for a diverse range of programming tasks, across different courses.
Table \ref{tab:coding-code-use} gives an overview of the themes we identified in our analysis, based on an existing taxonomy of tasks for which developers use GenAI tools \cite{tufano2024unveiling}.
There were several tasks from the original taxonomy that we did not find in the student responses, such as all subcategories of \lbl{process}, which contain activities related to the development process (e.g. release planning, automating the creation of commits, PRs, and issues).
From the \lbl{feature implementation} category we only kept \lbl{finding an API for a given task} and \lbl{implementing a new feature}. For the latter category, we identified two subcategories: generic uses of GenAI, and using GenAI to create boilerplate code, code templates, and other structures.
As expected in an educational setting, the \lbl{learning} category played a bigger role for students, therefore we modified the subcategories and identified a new subcategory for responses that talked about \lbl{understanding/starting a task}.

\begin{table}[tb]
\centering
\caption{Students' use of GenAI for coding tasks. Mentions for ProgCrs-BSc (cB), ProgReq-MSc (rM), ProgReq-BSc (rB), ProgOpt-MSc (oM), ProgOpt-BSc (oB). Subcategories marked with $\bigstar$ are new additions to the taxonomy. }
\label{tab:coding-code-use}
\sffamily
\small
\begin{tabular}{p{4cm} l rrrrrr r}
\hline
\textbf{Main category} & \textbf{Subcategory} & \multicolumn{5}{c}{\textbf{Mentions}}  & \textbf{Total}\\
 \hline
& & \textbf{cB} & \textbf{rM} & \textbf{rB} & \textbf{oM} & \textbf{oB} & \\
 
Feature implementation & Implementing features, generic use $\bigstar$ &  1 & 12 & 11 & 2 & 3 & 29 \\
                 & Implementing features, boilerplate code $\bigstar$  & 2 & 2 & 5 & 2 & 1 & 12 \\
                 & Finding an API for a given task  & 0 & 1 & 2 & 0 & 0 & 3 \\
Learning & Understanding task, head start $\bigstar$  & 3 & 5 & 2 & 0 & 0  & 10 \\
         & Program comprehension  & 3 & 5 & 3 & 0 & 0 & 11 \\
         & How to use a library/framework/tool  & 8 & 6 & 9 & 1 & 1 & 25 \\
         & Generating code examples  & 4 & 2 & 3 & 0 & 0  & 9 \\
         & Generic or unclear &  3 & 4 & 2 & 0 & 0  & 9 \\
Generating/manipulating data & -- &  0 & 1 & 2 & 0 & 0  & 3 \\
Software quality & Fixing &  16 & 35 & 20 & 2 & 4 & 77 \\
                 & Refactoring &  5 & 10 & 3 & 0 & 0  & 18 \\
                 & Testing &  0 & 0 & 5 & 0 & 1 & 6 \\
Development environment & -- &  0 & 1 & 0 & 0 & 0  & 1 \\
Documentation & Generating code comments  &  0 & 2 & 2 & 2 & 0 & 6 \\ \hline
Generic, or unclear use & -- &  1 & 6 & 7 & 2 & 0  & 16 \\
Did not use it &  & 28 & 22 & 51 & 5 & 9  & 115 \\

\hline
\end{tabular}
\end{table}

In the following subsections, we discuss the responses for the different types of courses. 

\subsubsection{Programming courses} 

The most prominent use of GenAI in programming courses regarding \lbl{software quality} was for debugging and fixing code. While most students simply mentioned debugging, three students explicated that they used debugging \quo{only when I had a bug i had tried to fix for a long time but couldn't succeed}. Two students added details on how they used the generated debugging help: \quo{I have used it for fixing a bug, but I did not use the code it provided and only used the idea}. Five students used GenAI for refactoring and no responses mentioned testing.

Several students used GenAI for \lbl{learning}. Asking for help on how to apply certain functions or libraries and on how certain programming concepts work was mentioned most. Few students generated code examples for exploration or asked GenAI to explain code they did not understand. Others used GenAI as a teaching assistant or for a head-start by either helping them in understanding the assignment or providing help on how to continue when completely stuck. 
Interestingly, the usage of GenAI for \lbl{feature implementation} seems to be scarce in programming courses. Only two students generated boilerplate code and two used GenAI for auto-completion and general code generation.
The rationale for this is expressed by a large number of students taking these courses: most students would prefer disallowing GenAI, as the very goal of the course is learning how to code: \quo{For a `basics' course like this you should be able to do everything in this course by yourself}.
A student from the first programming course said \quo{I think it should be disallowed as it might influence the understanding students develop for the course content}.

A notable observation for this course type is that we did not see any student quotes focusing on the impact of GenAI for their future careers, a pattern which was more dominated by answers from courses where programming is required, but that are not about actually learning to program. On the contrary, students taking core programming courses largely focused their answers on learning, and the need to disallow GenAI as it hinders learning. 

Several student also highlighted the need to use GenAI critically: \quo{students should always check what the answer given is, and whether it’s a good answer} and also the need to \quo{teach} students how to critically use it: \quo{learning how GenAI works, and incorporating it into the course}.

However, we still observed that the biggest group (29 students) stated that they have not used GenAI in their programming course at all. Three of these additionally indicated that they had used GenAI for other purposes and one student mentioned a likely use for debugging if a problem would have occurred.

\subsubsection{Courses in which programming is required}

In courses in which programming is required, but not a learning goal in itself, we see a shift in use regarding \lbl{feature implementation}. In several of these courses, both BSc and MSc students report using GenAI to generate boilerplate code (e.g. \quo{generate simple getters}, \quo{bulk work}, \quo{basic skeleton of test functions}), and other generic code generation uses (\quo{writing}, \quo{using a coding buddy}). Some students specifically mention they only use it for single lines, or \quo{max 5 lines of code}. The breadth of use cases also increases for these courses: we see new uses such as finding an API (\quo{finding out if some functions exist that i do not have enough knowledge of}), generating/manipulating data (\quo{fill database}), development environment tasks (\quo{get some Docker template}), documentation (\quo{add comments}), and testing, while all learning subcategories are also still relevant.

As an example, in a MSc course in the Applied Data Science program, in which students learn methods to extract, link and prepare data from medical systems, a student mentions \quo{writing a lot of code that just needs to be ran once anyway}, but also for \quo{throwing a database at it and letting it discover what's what}.
Generating test code was mostly mentioned by students taking a capstone software development course: \quo{to generate simple unit tests} and \quo{to help me use mocking in my tests}. \lbl{Refactoring} code was also a topic we noticed often in courses requiring programming, for example, we noticed several responses from students doing a course on data wrangling using GenAI to \quo{improve}, \quo{tidy up}, and \quo{restructuring} code.

Across all course types, students mention that it should be allowed for explaining and as an aid in understanding the material, but not for actually writing the entire code. In another context, a Data Mining course, a student gave an example on decision trees: \quo{GenAi could be used to explain Decision Trees in more detail, but copy and pasting a generated function to create a decision tree should not be used}.

Still, some students from these courses argue disallowing GenAI. A student noted that the reason they propose disallowing GenAI is that students should take responsibility for their own learning: \quo{For learning skills, people should be able to come up with results on their own. Don't give responsibility away to AI!}. Many students pointed out the need to disallow GenAI for tasks where understanding is central.

Other students made a connection with the \textit{course nature}. For courses that require students to develop more ``real-world'' and complex code, such as project-based and challenge-based learning courses where the programming task is given by an external client, then students argue that GenAI should be allowed: \quo{in real-world software development you can also use generative AI}, and \quo{[Using] GenAI is realistic for [one's] career}. 

Mostly students in this category argue that GenAI usage cannot be prevented (\quo{Forbidding it is futile}), and that it is impossible to check for teachers (\quo{it's impossible to police}). For this reason, some students think it should always be allowed, but some would actually prefer the opposite \quo{Ideally disallowed but there is no way to control it}.

\subsubsection{Courses in which programming is helpful}

For courses where programming is considered helpful but not required for its goals, students were more unanimously than the other categories in favor of allowing the use of GenAI.
We have also seen this in Figures \ref{fig:q19_level} \& \ref{fig:q19_course_type}, where MSc students were more in favor of always allowing it than BSc students for these courses, and have the same lower percentage of always disallowed.
They base their rationale on the fact that if GenAI is used for a task that is not central to the \textit{course goals}, then it should be allowed. 

As an example, for a MSc course on Crowd simulation, a student mentioned that GenAI was encouraged, and they even received extra points for it, under the condition it was disclosed. Other responses for these courses often mention that these courses are not about learning programming (\quo{the assignments are about figuring out strategies and not engineering} in the Multi-Agent systems course), so it should be allowed for these tasks.
Similarly, for a course on Software ecosystems security a student noted that \quo{The course is not about our ability to write clean code but rather the quality of our research}. For the course Personalisation for media, a student notes \quo{You're not teaching programming here, so just let people use AI}.

\section{Discussion}

\subsection{Answers to our RQs}

For RQ1 we investigate how GenAI tools are currently used by students across different types of programming courses.
In the analysis we distinguish between programming courses (where the fundamentals of programming are taught), courses where programming is a required component, and courses where programming is optional. Of the 269 responses on how students use GenAI tools (Q22), 115 responses indicated no use. Looking at the instances where GenAI was used, we found some differences between the different types of courses. Students relied more on GenAI for implementing features in courses where learning to program was not the main goal. This is also reflected in the answers of students when asked when they believe use of GenAI should be allowed, and can be connected to the differences we found in the acceptance of GenAI use between MSc and BSc students. MSc students found it more acceptable to use GenAI tools for programming assignments in their courses. A reason for this could be that these MSc courses were almost all programming required or programming optional courses, but not fundamental programming courses in which students were still learning the programming basics. The results for the BSc students also point towards this. The results show a trend where the use of GenAI tools is seen as more appropriate for courses where learning to program is not the goal but applying programming is. This finding aligns with the results of \citeauthor{margulieux2024self} \cite{margulieux2024self}, where the perceived usefulness of AI in an introductory programming course decreased over time. In the study, the findings indicate that students are still interested in learning, with some expressing skepticism toward the AI’s responses and recognizing when its outputs are inaccurate or unhelpful.

Our second research question (RQ2) focuses on how students' opinions on GenAI have changed over the course of an academic year.
Looking at the changes over time, we found a few significant results. Notably, one question asked students whether it is unethical to generate an entire solution to an assignment and hand this in. \citeauthor{prather2023robots} \cite{prather2023robots} reported a percentage of 95 on this question in the summer of 2023. In our survey, 99.2\% of respondents found this unethical in Period 1, but in Period 2 this percentage dropped to 92\%, and in Period 3 to 89.2\%. It seems as the year progressed, students found handing in an AI-generated assignment to be less problematic. From the current survey, we cannot conclude why this is the case, but it would be interesting to see if this trend continues into Period 4 and the new academic year. 

Some trends are visible both over time and across different types of courses. Our results show that students increasingly view GenAI as important help-seeking tools when encountering difficulties. Although online searches are still preferred over other methods, over the course of our study GenAI has surpassed the teachers and TA's. This aligns with the results of \citeauthor{hou2024effects} \cite{hou2024effects}. In our study, we observe a difference in type and levels of courses. MSc students rank GenAI tools as better support tools than BSc students. Between different types of BSc courses, we also found that students in introductory programming courses prioritize GenAI tools less as support tool than in other types of BSc courses. 

While we did not specifically analyze gender differences in GenAI usage, other studies have highlighted variations between different demographic groups. For instance, \citeauthor{margulieux2024self} \cite{margulieux2024self} found that women tended to incorporate AI assistance later in the problem-solving process. Similarly, \citeauthor{gooch2024exploring} \cite{gooch2024exploring} reported that male students were more likely to have their assessments flagged for potential AI-generated content than their female counterparts. These findings suggest potential gender differences in GenAI use.

\subsection{Limitations}
First, this study was conducted in a single institute in the Netherlands. Policies as stated by the institute and university could have influenced the perceptions and behavior of students. It would be interesting to replicate this longitudinal study at other universities and in other countries.

Secondly, the students who responded to the survey might have been interested in GenAI more than the average student, leading them to feel the need to voice their opinions and experiences. This could impact the results in multiple ways. For example, these students might have stronger opinions, either more positive or more negative, about GenAI. This could affect how representative our findings are of the broader student population.

Third, students' reports may not accurately reflect reality. They might be reluctant to admit using GenAI for unapproved purposes in a course due to fear of retaliation, even though the responses were treated confidentially.

Fourth, students could skip the first part of the survey if they had already completed that part. Therefore, we cannot track individuals' opinions over time. We can only investigate the overall changes in opinion of the respondents as a group. Because we invited students from the same study programs, we do not believe this poses a threat for the interpretation of our results.

Finally, the distribution of the different types of courses over the periods was uneven. The core programming courses were mainly represented in Period 1, while the other periods contained more responses for programming required and programming optional courses. 

\section{Conclusion and Future Work}
In this study, we aimed to explore the current use of GenAI tools in various programming courses and how students' opinions on GenAI evolve over time. To achieve this, we conducted a survey among students from the Department of Information and Computing Sciences at a large European research university at three different moments during the 2023--2024 academic year. The survey, based on a previously conducted international study, was divided into two sections: general opinions on the use of GenAI and course-specific questions. We received 249 responses for the first part and 340 for the second part. Our analysis used both quantitative and qualitative methods, focusing on trends over time and categorizing courses into five distinct types.

Our results show several key trends and insights:

\begin{itemize}
    \item While the use of GenAI tools has increased overall, some concerns about their use persist. Our results show an increase in the number of students who consider it ethical to autogenerate a complete solution for an assignment and submit it without understanding it.
    \item GenAI is gaining preference as a help-seeking strategy, with MSc students ranking it higher than BSc students.
    \item MSc students are more inclined to allow GenAI usage in their assignments compared to BSc students, who exhibit a consistently low and stable level of acceptance over time.
    \item Students in core programming courses emphasize the importance of learning and the need to disallow GenAI, as they believe it hinders their learning process. In other courses that use programming as a tool, student place more emphasis on GenAI being part the future, therefore justifying its use.
    \item Students rely more on GenAI for implementing features in courses where programming is not the main focus.
    \item Many students are still not using GenAI for coding-related tasks.
\end{itemize}

A recent column argued that students should be consulted when designing policies for GenAI use in educational settings.\footnote{Need a policy for using ChatGPT in the classroom? Try asking students. Maja Zonjić. June 2024. \url{https://www.nature.com/articles/d41586-024-01691-4}}
Our survey results, complementing other surveys discussed earlier, show the evolving landscape of insights, concerns, and ideas students have about this topic.

Moving forward, we plan to extend our analysis by incorporating additional data. At certain intervals, we will analyze the collected data to determine whether the observed trends persist or if new ones emerge. Moreover, we aim to conduct a more thorough analysis. For example, correlating demographics and programming experience with the use of GenAI to identify any significant differences. Furthermore, we would like to expand this study to include data from multiple universities within the Netherlands and throughout Europe.

\begin{acks}
We would like to thank all the students that participated in the survey, the teachers that helped distributing the survey, and Utrecht University's Faculty Education Incentive Fund for funding the project in which we conducted this study.
\end{acks}


\bibliographystyle{ACM-Reference-Format}
\bibliography{refs}

\clearpage
\appendix
\section*{Survey questions}
\label{survey}

\subsection*{Demographics part}
\begin{enumerate}

\item{
\textbf{How do you describe yourself?}
\begin{itemize}
    \item[\Circle] Male
    \item[\Circle] Female
    \item[\Circle] Non-binary/third gender  
    \item[\Circle] Prefer to self-describe: 
    \item[\Circle] Prefer not to say  
\end{itemize}
}

\item{\textbf{Were you born in the Netherlands?}
\begin{itemize}
    \item[\Circle] Yes
    \item[\Circle] No
\end{itemize}    
}

\item{
\textbf{What is your level of study:}
\begin{itemize}
    \item[\Circle] Bachelor year 1  
    \item[\Circle] Bachelor year 2
    \item[\Circle] Bachelor year 3
    \item[\Circle] Bachelor year 4+  
    \item[\Circle] Masters student  
    \item[\Circle] PhD student   
\end{itemize}
}

\item{
\textbf{Select the program you're studying}
\begin{itemize}
    \item[\Circle] BSc Information Science
    \item[\Circle] BSc Computer Science
    \item[\Circle] BSc Game Technology
    \item[\Circle] MSc Applied Data Science 
    \item[\Circle] MSc Artificial Intelligence
    \item[\Circle] MSc Business Informatics
    \item[\Circle] MSc Computing Science   
    \item[\Circle] MSc Data Science  
    \item[\Circle] MSc Game and Media Technology  
    \item[\Circle] MSc Human Computer Interaction  
    \item[\Circle] Other:
\end{itemize}
}

\item{\textbf{How many courses with a programming component have you completed?} 
\begin{itemize}
    \item[\Circle] 4 or fewer
    \item[\Circle] 5 to 9
    \item[\Circle] 10 to 14
    \item[\Circle] 15 to 19 
    \item[\Circle] 20 to 29
    \item[\Circle] 30 or more
\end{itemize}
}

\item{\textbf{Please estimate, on a scale from 1 to 10, your proficiency in the following programming-related skills/topics:}
\begin{itemize}
    \item Programming in general
    \item Object-oriented programming (C\#, Java, ...) 	 
    \item Functional programming (Haskell, Clojure, ...) 	 
    \item Web development 	
    \item Database technologies  
\end{itemize}
}

\item{\textbf{Please estimate, on the scale from 1 to 10, how important you think programming is going to be for your future career.}
}

\end{enumerate}

\subsection*{General part}
\begin{enumerate}

\setcounter{enumi}{7}
\item{
\textbf{Rate your agreement with the following statements:} (5-point Likert)
\begin{itemize}
    \item I regularly use GenAI tools when working with text (e.g.: writing emails, reports, summaries)
    \item I regularly use GenAI tools when working with code (e.g.: generating code or explanations, writing programs, debugging, ...)
    \item I regularly use GenAI tools when working with images (e.g.: generating new pictures, ...)
\end{itemize}
}

\item{
\textbf{Rate your agreement with the following statements:} (5-point Likert)
\begin{itemize}
    \item I expect to use GenAI tools increasingly in my learning practices in the future
    \item Using GenAI tools frequently to generate code is harmful for my learning of programming
    \item GenAI tools can provide guidance for coursework as effectively as human teachers
    \item GenAI tools will replace human teachers in the future
    \item Students must be taught how to use GenAI tools well for their future careers
\end{itemize}
}

\item{
\textbf{Rate your agreement with the following statements:} (Likert)
\begin{itemize}
    \item The policies of [university] are clear regarding what is allowed and what is not allowed in terms of using GenAI tools
    \item The policies of the [department] are clear regarding what is allowed and what is not allowed in terms of using GenAI tools
    \item The policies in the courses I have been taking are clear regarding what is allowed and what is not allowed in terms of using GenAI tools
    \item There should be no restrictions on the use of GenAI tools in coursework
\end{itemize}
}

\item{
\textbf{To what extent do you think students at your school are using GenAI tools in ways that your instructors would not approve of?}
\begin{itemize}
    \item[\Circle] Almost everyone
    \item[\Circle] Many
    \item[\Circle] Some
    \item[\Circle] A few
    \item[\Circle] Almost none
\end{itemize}
}

\item{
\textbf{In the absence of an explicit course policy on the use of GenAI tools, which of the following do you consider as NOT ethical? (Select all that apply):}

\begin{itemize}
    \item[\Square] It is unethical to auto-generate a solution for the whole assignment (or a large portion of it) and submitting it without understanding it.

    \item[\Square] It is unethical to auto-generate a solution for the whole assignment (or a large portion of it) and submitting it after reading it and completely understanding it.
    
    \item[\Square] It is unethical to auto-generate a solution even for small parts of the assignment.
    
    \item[\Square] It is unethical to use GenAI tools to "explain" to you (step-by-step) how to solve the problem.
    
    \item[\Square] It is unethical to provide your code to GenAI tools and ask them to help you fix a bug.
    
    \item[\Square] It is unethical to ask GenAI tools to comment, tidy and improve the style of your code.
    
    \item[\Square] It is unethical to write the solution in a programming language (other than the one used in the course) and asking GenAI tools to translate it for you to the language of the course (and then submitting the translated code).
    
\end{itemize}
}

\item{
\textbf{Rate your agreement with the following statements:} (5-point Likert)
\begin{itemize}
    \item GenAI tools will negatively impact my future job prospects
    \item GenAI tools will harm the development of generic or transferable skills such as teamwork, problem-solving, and leadership
    \item I am concerned that I will become over reliant on GenAI tools
    \item I trust the code written by GenAI tools more than the code I write
    \item My instructors can detect code that was written by GenAI tools
    \item My instructors actively check for unauthorized use of GenAI tools
    \item If my instructors disallow GenAI tools, it's ok to use them to generate code if I understand the code. It is unethical only if I copy w/o understanding
    \item If everyone in class is using GenAI tools, but it is against the rules to use them, then I would still use them
\end{itemize}
}

\item{
\textbf{Describe the effects you think GenAI tools will have on your prospects for future employment:} (text entry)
}

\item{
\textbf{What are your views on the allowed usage of GenAI tools in coursework/exams?} (text entry)
}

\item{
\textbf{Which of the following courses did you follow in block 1/2/3 of this academic year (2023-24)?} (list of courses, multiple selections possible)
}

\end{enumerate}

\subsection*{Course-specific part (presented for each course selected in the previous question)}

Please answer the next set of questions with the course [name selected course] in mind.

\begin{enumerate}

\setcounter{enumi}{16}

\item{
\textbf{When you had a question regarding the material you were studying in this course or got stuck on a problem, in what order did you do the following? Please order the options below.}
\begin{itemize} 
    \item Ask using GenAI tools
    \item Ask on the course discussion forum
    \item Search online (e.g. with Google)
    \item Ask a friend or classmate
    \item Ask on online forums like StackOverflow
    \item Ask the course instructor/TA
\end{itemize}
}

\item{
\textbf{After generating code using GenAI tools, in my assignments for this course, I mostly did the following:}

\begin{itemize}
    \item[\Circle] Not applicable (I have not used GenAI tools to generate code)
    \item[\Circle] Use the code immediately.
    \item[\Circle] Skim through the code briefly to make sure that it looks correct.
    \item[\Circle] Read it carefully (with skepticism) to ensure that it is correct.
    \item[\Circle] Read it carefully (with skepticism) and also write code to test it.
\end{itemize}
}

\item{
\textbf{For programming assignments in this course, I believe GenAI should be:}

\begin{itemize}
    \item[\Circle] Always allowed
    \item[\Circle] Allowed in some assignments, disallowed in others (based on the assignment type, course level, etc.)
    \item[\Circle] Always disallowed
\end{itemize}
}

\item{
\textbf{Can you elaborate on when you believe GenAI should be allowed or disallowed in this course?} (text entry)
}

\item{
\textbf{Describe the ways you currently use GenAI tools in this course for text generation (e.g.: writing reports, summaries, etc.)} (text entry)
}

\item{
\textbf{Describe the ways you currently use GenAI tools in this course for code generation (e.g.: debugging, writing, etc.)} (text entry)
}

\end{enumerate}

\newpage
\section*{Course list}

\begin{table}[h!]
\centering
\caption{Course List - Part 1}
\begin{tabular}{|c|l|c|c|}
\hline
\textbf{Period} & \textbf{Name} & \textbf{Course Type} & \textbf{Level} \\ \hline
3 & Advanced functional programming & ProgCrs & MSc \\ \hline
2 & Advanced graphics & ProgReq & MSc \\ \hline
2 & Advanced machine learning & ProgReq & MSc \\ \hline
1 & AI for game technology & ProgReq & MSc \\ \hline
1 & Algorithms for decision support & ProgReq & MSc \\ \hline
3 & Algoritmiek (Algorithms) & ProgReq & BSc \\ \hline
2 & Applications of machine learning & ProgReq & BSc \\ \hline
1 & Beeldverwerking (Image processing) & ProgReq & BSc \\ \hline
3 & Causal inference methods for policy evaluation & ProgOpt & MSc \\ \hline
1 & Cloud and Edge Computing & ProgReq & MSc \\ \hline
2 & Cognitive Modeling & ProgReq & MSc \\ \hline
2 & Computationeel denken (Computational thinking) & ProgCrs & BSc \\ \hline
2 & Computationele intelligentie (Computational intelligence) & ProgReq & BSc \\ \hline
2 & Computer animation & ProgReq & MSc \\ \hline
2 & Concepts of programming language design & ProgReq & MSc \\ \hline
2 & Concurrency & ProgReq & BSc \\ \hline
2 & Crowd simulation & ProgOpt & MSc \\ \hline
1 & Data analytics & ProgReq & BSc \\ \hline
1 & Data mining & ProgReq & MSc \\ \hline
1 & Data science and society & ProgReq & BSc \\ \hline
1 & Data Wrangling & ProgReq & MSc \\ \hline
3 & Databases & ProgReq & BSc \\ \hline
1 & Datamodelleren (Data modelling) & ProgReq & BSc \\ \hline
1 & Functioneel programmeren (Functional programming) & ProgCrs & BSc \\ \hline
3 & Game physics & ProgReq & MSc \\ \hline
3 & Game-ontwerp (Game design) & NoProg & BSc \\ \hline
1 & Gameprogrammeren (Games programming) & ProgCrs & BSc \\ \hline
3 & Human network analysis & ProgReq & MSc \\ \hline
1 & Imperatief programmeren (Imperative programming) & ProgCrs & BSc \\ \hline
2 & Informatica introductieproject (CS Introduction project) & ProgReq & BSc \\ \hline
1, 2, 3 & Informatica softwareproject (CS Software project) & ProgReq & BSc \\ \hline
2 & Informatie-uitwisseling (Information exchange) & ProgReq & BSc \\ \hline
3 & Intelligente systemen (Intelligent systems) & ProgReq & BSc \\ \hline
3 & Interactie-technologie (Interaction technology) & ProgReq & BSc \\ \hline
3 & Introduction to complex systems & ProgReq & BSc \\ \hline
2 & Knowledge and Data Engineering & ProgReq & MSc \\ \hline
\end{tabular}
\label{table:courselist_part1}
\end{table}

\begin{table}[h!]
\centering
\caption{Course List - Part 2}
\begin{tabular}{|c|l|c|c|}
\hline
\textbf{Period} & \textbf{Name} & \textbf{Course Type} & \textbf{Level} \\ \hline
3 & Knowledge-intensive process analysis & ProgReq & BSc \\ \hline
1 & Machine learning & ProgReq & BSc \\ \hline
1 & Methods in AI research & ProgReq & MSc \\ \hline
1 & Modelleren en systeemontwikkeling (Modelling and systems development) & ProgReq & BSc \\ \hline
1 & Motion and manipulation & ProgReq & MSc \\ \hline
3 & Multi-agent systems & ProgOpt & MSc \\ \hline
3 & Ontwerpen van interactieve systemen (Designing interactive systems) & NoProg & BSc \\ \hline
2 & Pattern Recognition and Deep Learning & ProgReq & MSc \\ \hline
3 & Personalisation for (public) media & ProgOpt & MSc \\ \hline
3 & Philosophy of AI & NoProg & MSc \\ \hline
1 & Probabilistic reasoning & ProgOpt & MSc \\ \hline
3 & Programmeren in Python, Programming with data, and Computational thinking & ProgCrs & BSc \\ \hline
2 & Requirements engineering & ProgOpt & MSc \\ \hline
2 & Scientific research methods & NoProg & MSc \\ \hline
2 & Small project Game and Media Technology & ProgReq & MSc \\ \hline
3 & Software ecosystems security & ProgOpt & MSc \\ \hline
2 & Sound and music technology & ProgOpt & MSc \\ \hline
2 & Systeemontwikkelingsmethoden (System development methods) & ProgReq & BSc \\ \hline
3 & Transformers: Applications in Language and Communication & ProgReq & MSc \\ \hline
3 & Usability engineering \& user experience & ProgOpt & BSc \\ \hline
3 & Using data from routine care & ProgReq & MSc \\ \hline
3 & Webtechnologie (Web technologies) & ProgReq & BSc \\ \hline
1 & Wetenschappelijke onderzoeksmethoden (Scientific research methods) & ProgReq & BSc \\ \hline
\end{tabular}
\label{table:courselist_part2}
\end{table}

\end{document}